\documentclass[twocolumn,prd,groupedaddress,nofootinbib,showpacs]{revtex4}
\usepackage{amsmath}
\usepackage{latexsym}
\usepackage{amssymb}
\usepackage{graphicx}
\usepackage{dcolumn}
\usepackage{bm}
\usepackage{verbatim}
\usepackage{subfigure}

\begin{document}

\title{Discussion on massive gravitons and propagating torsion in
arbitrary dimensions}
\author{C. A. Hernaski\footnote{carlos@cbpf.br}, A. A. Vargas-Paredes\footnote{alfredov@cbpf.br}, J. A. Helay\"{e}l-Neto\footnote{helayel@cbpf.br}}
\affiliation{Centro Brasileiro de Pesquisas F\'{\i}sicas, Rua Dr. Xavier Sigaud 150, Urca,%
\\
Rio de Janeiro, Brazil, CEP 22290-180}
%\date{\today}

\begin{abstract}
In this paper, we reassess a particular $R^{2}$-type gravity action in D
dimensions, recently studied by Nakasone and Oda, now taking torsion effects
into account. Considering that the vielbein and the spin connection carry
independent propagating degrees of freedom, we conclude that ghosts and
tachyons are absent only if torsion is nonpropagating, and we also conclude
that there is no room for massive gravitons. To include these excitations,
we understand how to enlarge Nakasone-Oda's model by means of explicit
torsion terms in the action and we discuss the unitarity of the enlarged
model for arbitrary dimensions.
\end{abstract}

\maketitle

\section{Introduction}

Massive gravity has been an issue of particular interest since the early
days of quantum gravity. More recently, in connection with models based on
brane-world scenarios, the discussion of \ massive gravitons is drawing a
great deal of attention, in view of the possibility of their production at
LHC and the feasibility of detection of quantum gravity effects at the TeV
scale \cite{lhc1}-\cite{nieto}. In the framework of branes, the graviton
acquires mass via a spontaneous breakdown of general coordinate
reparametrization symmetry \cite{Zurab}. However, as it is usual in all
Higgs-type mechanisms, a nonvanishing vacuum expectation value for an extra
scalar field is needed in the description. There is also an alternative way
to generate mass in three dimensions, as proposed by Jackiw, Deser and
Templeton \cite{jackiw}. There, a topological parity-violating term is added
to the Einstein-Hilbert gravity Lagrangian in order to describe a massive
graviton. The final theory is also unitary.

In this context, we asked if it is possible to build up a unitarity and
parity-preserving model that generates mass for the graviton without the
need of an extra field. Bergshoeff, Hohm and Townsend obtain such a model for $D=3$ 
\cite{ma} by considering a nonlinear theory that is equivalent to the
Pauli-Fierz model at the linear level.

In a very recent paper, M. Nakasone and I. Oda \cite{ja} have shown that a
particular $R^{2}$-type action in three dimensions is equivalent to the
massive Pauli-Fierz gravity at the linear level, as it has been proposed in 
\cite{ma}; moreover, they also describe how, only in three dimensions, there
is no ghost, so that the model preserves unitarity. In fact, the question of
unitarity in massive gravity theories is a topic of special relevance in the
literature \cite{jackiw}-\cite{thooft}.

Besides these considerations, massive gravity is of interest by itself. For
example, the work of Ref. \cite{Witten1} has pointed out the relevance of
three-dimensional gravity in connection with Conformal Field Theories (CFT) theories \cite{LSS}.
Three-dimensional gravity has no local degrees of freedom. The Riemann
tensor has the same number of components as the Ricci tensor, which means
that all solutions in these theories are trivial, with the exception of those
that consider topological effects. However, the situation might change if we
consider massive spin-2 propagating modes in three dimensions. This is
because the Poincar\'{e} group representations of massive particles in three
dimensions and massless particles in four dimensions are described by the
same little group, $SO\left( 2\right) $, having both two types of helicities 
$\pm 2\ $\cite{ma}. We do not however discuss these interesting points in
the present paper.

Specifically, we investigate if there is a possible generalization\ of the
results of \cite{ja}, whenever we have propagating torsion in any dimension. We work with the
vielbein and the spin connection as independent fields. Our viewpoint is
that this is a more fundamental approach to gravitation, since it is based
on the fundamental ideas of the Yang-Mills approach \cite{utiyama}, \cite%
{kibble}. As it shall become clear in the sequel, we conclude that explicit
terms in the torsion field are needed in order to describe a propagating
massive graviton. 

We also analyze the unitarity of the model,
and for this we consider the most general parity-preserving Lagrangian
without higher derivatives in D dimensions. We obtain a certain number of
unitary Lagrangians that yield a propagating massive graviton and compare
them with those Lagrangians found by Sezgin and Nieuwenhuizen \cite{bu}, in case we reduce our results to D=4. As we consider only quadratic terms in the curvature and torsion in the Lagrangian, by virtue of the Gauss-Bonet theorem, there is a redundant term among the possibilities for D=4. But, for $D\neq4$, this term must be considered, since the Gauss-Bonet theorem does not involve quadratic terms for $D\neq4$. So, structurally, this is the important difference from the Lagrangian considered in \cite{bu}. The outcome is that we find a set of
Lagrangians with a massive graviton that, in the particular case of D=4, reproduce those studied in 
\cite{bu}. However, we should mention that we have no intention to reproduce and repeat all the results of \cite{bu}, where there is an exhaustive and complete analysis. We carry out our discussion in D dimensions and, for $D=4$, we shall be pointing out the cases that correspond to intersections with situations contemplated in \cite{bu}.

Our paper is organized according to the following outline: in Sec. II, we present the model, our conventions and obtain the propagators of the corresponding modes. Next, Sec. III tackles the question of how to introduce massive gravitons by enlarging our initial model. In Sec. IV our aim is to analyze the existence of tachyons or ghost modes in the massive and nonmassive sectors. Finally, in Sec. V, we set up our concluding remarks. In the Appendix, we collect the whole set of spin operators that appear in our treatment.

\section{Description of the Model}

In order to investigate the changes that occur when torsion propagates, we
start off by considering the same Lagrangian as the one analyzed by Nakasone
and Oda in Ref.~\cite{ja}, with the exception \ that we consider here the
right sign of the Einstein-Hilbert term. In the models of Refs. \cite{ma}, 
\cite{ja}, the opposite sign is essential for the reduction of the
Lagrangian to the Pauli-Fierz model. However, as shown in~\cite{ja}, this
reduction is possible only in three dimensions. This can be seen by noticing
that, in three dimensions, the Einstein-Hilbert Lagrangian does not have any
propagation mode, whereas in dimensions higher than three it does propagate
a unitary massless mode. With the "wrong" sign, the model necessarily
displays ghosts in the spectrum. Therefore, our starting point is the
Lagrangian below:%
\begin{multline}
$$\cal{L}$$_{R}=e\left( -\frac{1}{\kappa ^{2}}R+\alpha R^{2}+\beta R_{\mu a}R^{\mu
a}\right)   \label{lagrangian} \\
+e\gamma \left( R_{\mu \nu ab}R^{\mu \nu ab}-4R_{\mu a}R^{\mu
a}+R^{2}\right) ,
\end{multline}%
\qquad where $\alpha $, $\beta $, $\gamma $, are arbitrary dimensionless
constants and $e$ is the determinant of the vielbein. In the work of \\\\Ref. 
\cite{ja}, their values are set to be

\begin{equation}
\alpha=-\frac{D}{4-D}\beta,\ \ \ \gamma=0.\ 
\end{equation}
We do not adopt these choices here, because now the Lagrangian (\ref%
{lagrangian})\ contains $R,\ R_{\mu}^{~~a},\ R_{\mu\nu}^{~~~~ab}$ with the
vielbein, $e_{\mu}^{a},$ and the spin connection, $\omega_{\mu}^{~~ab},$
taken as independent fields. We must analyze if this yields a consistent
quantum theory as far as unitarity and causality are concerned. Our
conventions are

\begin{subequations}
\begin{gather}
R_{\mu\nu}^{~~~\
ab}=\partial_{\mu}\omega_{\nu}^{~~ab}-\partial_{\nu}\omega_{\mu}^{~~ab}+%
\omega_{\mu~~c}^{~~a}\omega_{\nu}^{~~cb}-\omega_{\nu
~~c}^{~~a}\omega_{\mu}^{~~cb}, \\
R_{\mu}^{~~a}=e_{b}^{\nu}R_{\mu\nu}^{~~~\ ab}, \\
R=e_{a}^{\mu}e_{b}^{\nu}R_{\mu\nu}^{~~~\ ab}, \\
\eta_{ab}=\left( 1,-1,-1,-1\right) ,
\end{gather}
where the Greek indices refer to the world manifold and the Latin ones stand
for the frame indices.

In the following, we shall consider fluctuations of the fundamental fields
in order to set up the quantum theory:

\end{subequations}
\begin{subequations}
\begin{align}
e_{\mu}^{a} & =\delta_{\mu}^{a}+\widetilde{e}_{\mu}^{~~a}\ , \\
\omega_{\mu}^{~ab} & =\widetilde{\omega}_{\mu}^{~ab}.
\end{align}
We also define the $\phi$ and $\chi$ fields as 
\end{subequations}
\begin{subequations}
\begin{align}
\phi_{ab} & =\frac{1}{2}\left( \widetilde{e}_{ab}+\widetilde{e}_{ba}\right)
,\  \\
\ \chi_{ab} & =\frac{1}{2}\left( \widetilde{e}_{ab}-\widetilde{e}%
_{ba}\right) .
\end{align}

The Lagrangian, up to second-order terms in the quantum fluctuations, can be
written as

\end{subequations}
\begin{equation}
\left( \cal{L}_{R}\right) _{2}=\displaystyle\sum_{\alpha, \beta}\psi _{\alpha
}O_{\alpha \beta }\lambda _{\beta },
\end{equation}%
where $\psi _{\alpha }$, $\lambda _{\alpha }$ carry the 40 components $%
\left( \phi _{ab},\chi _{ab},\omega _{abc}\right) $.

In order to investigate the spectrum of our model, we work with a complete
set of spin projector operators for a conserved parity model describing a
rank-3 antisymmetric tensor in two indices and a rank-2 tensor. With the help of these operators, given in the Appendix
we split the bilinear piece of the Lagrangian as:

\begin{equation}
\left( \cal{L}_{R}\right) _{2}=\displaystyle\sum_{\alpha,\beta,ij,J^{P}}%
\psi_{\alpha}a_{ij}^{\psi\lambda}\left( J^{P}\right) P_{ij}^{\psi\lambda
}\left( J^{P}\right) _{\alpha\beta}\lambda_{\beta}.
\end{equation}

Here, we adopt the conventions of Ref. \cite{bu}. The diagonal operators, $%
P_{ii}^{\Psi\Psi}\left( J^{P}\right) $, are projectors in the spin $\left(
J\right) $ and parity $\left( P\right) $ sectors of the field $\Psi$. The
off-diagonal operators $(i\neq j)$ implement mappings inside the spin/parity
subspace. These operators form a basis with a completeness relationship:

\begin{subequations}
\begin{align}
P_{ij}^{\Sigma\Pi}\left( J^{P}\right) _{\alpha\beta}P_{kl}^{\Lambda\Xi
}\left( I^{Q}\right) _{\beta\gamma} &
=\delta^{PQ}\delta^{\Pi\Xi}\delta_{jk}\delta_{IJ}P_{il}^{\Sigma\Xi}\left(
J^{P}\right) _{\alpha\gamma }, \\
\displaystyle\sum_{i,\alpha,\beta,J^{P}}P_{ii}\left( J^{P}\right)
_{\alpha\beta} & =\delta_{\alpha\beta}.
\end{align}

The $a_{ij}\left( J^{P}\right) $ coefficient matrices, representing the
contribution to the spin $\left( J\right) $ and parity $\left( P\right) $,
are given by

\end{subequations}
\begin{equation*}
\end{equation*}

\begin{widetext} 
\begin{equation}
 a_{ij}\left( 2^{+}\right) =%
\begin{array}{c}
\omega  \\ 
\phi 
\end{array}%
\overset{%
\begin{array}{cc}
\omega \ \ \ \ \  & \ \ \ \ \ \ \ \ \ \phi 
\end{array}%
}{\left( 
\begin{array}{cc}
-\frac{1}{2\kappa ^{2}}+\frac{1}{2}\beta p^{2} & \ \ \ \ i\sqrt{p^{2}}\frac{1%
}{\sqrt{2}\kappa ^{2}} \\ 
-i\sqrt{p^{2}}\frac{1}{\sqrt{2}\kappa ^{2}} & 0%
\end{array}%
\right) ,}
\end{equation}

\begin{equation}
a\left( 2^{-}\right) =-\frac{1}{2\kappa^{2}}+2\gamma p^{2},
\end{equation}

\begin{equation}
a_{ij}\left( 1^{+}\right) =%
\begin{array}{c}
\omega  \\ 
\omega  \\ 
\chi 
\end{array}%
\overset{%
\begin{array}{ccc}
\omega \ \ \ \ \  & \ \ \ \ \ \ \ \ \omega \  & \ \ \ \ \ \ \ \ \ \ \ \chi \ 
\end{array}%
}{\left( 
\begin{array}{ccc}
\frac{1}{2\kappa ^{2}}+\frac{1}{2}\beta p^{2} & \ \ \ \ -\frac{1}{\sqrt{2}}%
\frac{1}{\kappa ^{2}} & \ \ \ i\sqrt{p^{2}}\frac{1}{\sqrt{2}\kappa ^{2}} \\ 
-\frac{1}{\sqrt{2}}\frac{1}{\kappa ^{2}} & \ \ 0 & 0 \\ 
i\sqrt{p^{2}}\frac{1}{\sqrt{2}\kappa ^{2}} & \ \ 0 & 0%
\end{array}%
\right) },
\end{equation}

\begin{equation}
{a_{ij}\left( 1^{-}\right) =\overset{%
\begin{array}{cccc}
\ \ \ \ \ \ \ \ \omega\ \ \ \ \ \ \ \ \ \  & \ \ \ \ \ \ \ \ \omega\ \ \ \ \
\  & \ \ \ \ \ \ \ \ \ \ \ \ \ \phi & \ \ \ \ \ \ \ \ \ \ \ \ \ \ \ \ \ \ \
\ \ \ \ \ \ \ \chi \ 
\end{array}
}{%
\begin{array}{c}
\omega \\ 
\omega \\ 
\phi \\ 
\chi%
\end{array}
\left( 
\begin{array}{cccc}
\begin{array}{c}
\frac{\left( D-2\right) }{2}\beta p^{2}~+~(D-3)\frac{1}{2\kappa^{2}} \\ 
-2\left( D-3\right) \gamma p^{2}%
\end{array}
& -\frac{\left( D-2\right) ^{1/2}}{2}\frac{1}{\kappa^{2}} & -\frac {1}{%
2\kappa^{2}}i\sqrt{p^{2}}\left( D-2\right) ^{1/2} & -\frac{1}{2\kappa^{2}}i%
\sqrt{p^{2}}\left( D-2\right) ^{1/2} \\ 
-\frac{\left( D-2\right) ^{1/2}}{2}\frac{1}{\kappa^{2}} & 0 & 0 & 0 \\ 
\frac{1}{2\kappa^{2}}i\sqrt{p^{2}}\left( D-2\right) ^{1/2} & 0 & 0 & 0 \\ 
\frac{1}{2\kappa^{2}}i\sqrt{p^{2}}\left( D-2\right) ^{1/2} & 0 & 0 & 0%
\end{array}
\right) }, }
\end{equation}

\begin{equation}
a_{ij}\left( 0^{+}\right) =%
\begin{array}{c}
\omega  \\ 
\phi  \\ 
\phi 
\end{array}%
\overset{%
\begin{array}{ccc}
\ \ \ \ \ \ \ \ \ \ \ \ \ \ \ \ \ \ \ \ \ \ \ \ \ \omega \ \ \ \ \ \ \ \ \ \
\ \ \ \ \ \ \ \ \ \ \ \ \ \  & \ \phi  & \ ~~\ \ \ \ \phi 
\end{array}%
}{\left( 
\begin{array}{ccc}
\frac{D}{2}\beta p^{2}-\frac{1}{\kappa ^{2}}\left( 1-\frac{D}{2}\right)
+2\left( D-1\right) \alpha p^{2} & \ \ \ -i\sqrt{p^{2}}\frac{\left(
D-2\right) }{\sqrt{2}\kappa ^{2}} & \ \ 0 \\ 
i\sqrt{p^{2}}\frac{\left( D-2\right) }{\sqrt{2}\kappa ^{2}} & 0 & 0 \\ 
0 & 0 & 0%
\end{array}%
\right) },
\end{equation}

\begin{equation}
a\left( 0^{-}\right) =\frac{1}{\kappa ^{2}}+2\gamma p^{2}.
\end{equation}

\end{widetext}

As it can be readily seen, the matrices for the spins $J^{P}=(1^{\pm},0^{+})$
are degenerate; this reflects the fact that there are some local invariances
in our Lagrangian. We already expected this, since our model is invariant
under linearized general coordinates and local Lorentz transformations. If
these matrices were invertible, propagators saturated with the external
sources could be written as

\begin{equation}
\Pi=i\displaystyle\sum S_{\alpha}^{\ast}a_{ij}^{-1\psi\phi}P_{ij}^{\psi\phi}\left(
J^{P}\right) _{\alpha\beta}S_{\beta},  \label{propagator}
\end{equation}
with the S$_{\alpha}$'s being the sources $\tau_{abc},\Sigma_{\left(
ab\right) ,}\Sigma_{\left[ ab\right] }$ for the spin connection, the
symmetric and the antisymmetric parts of the vielbein, respectively. These
sources satisfy suitable constraints imposed by the gauge invariances of the
action. In addition to the symmetries shared by the sources and fields, they
are conserved. They then satisfy

\begin{equation}
\partial_{a}\tau^{abc}\equiv\partial_{a}\Sigma^{\left( ab\right)
}\equiv\partial_{a}\Sigma^{\left[ ab\right] }=0.
\label{sources conservation}
\end{equation}

In the present case of degenerate matrices, the correct propagator is
obtained by taking the inverse of the largest nondegenerate submatrix and
saturating it with the conserved sources. Since these sources are conserved,
the resulting propagator is gauge invariant, as shown in \cite{bd}. The
nondegenerate matrices are given by%
\begin{equation*}
\end{equation*}%
\begin{widetext}

\begin{equation}
a_{ij}\left( 2^{+}\right) =\left( 
\begin{array}{cc}
-\frac{1}{2\kappa^{2}}+\frac{1}{2}\beta p^{2} & i\sqrt{p^{2}}\frac{1}{\sqrt {%
2}\kappa^{2}} \\ 
-i\sqrt{p^{2}}\frac{1}{\sqrt{2}\kappa^{2}} & 0%
\end{array}
\right) ,  \label{1}
\end{equation}

\begin{equation}
b_{ij}\left( 1^{+}\right) =\left( 
\begin{array}{cc}
\frac{1}{2\kappa^{2}}+\frac{1}{2}\beta p^{2} & -\frac{1}{\sqrt{2}}\frac {1}{%
\kappa^{2}} \\ 
-\frac{1}{\sqrt{2}}\frac{1}{\kappa^{2}} & 0%
\end{array}
\right) ,
\end{equation}

\begin{equation}
b_{ij}\left( 1^{-}\right) =\left( 
\begin{array}{cc}
\begin{array}{c}
\frac{\left( D-2\right) }{2}\beta p^{2}~+~\frac{D-3}{2\kappa^{2}} \\ 
-2\left( D-3\right) \gamma p^{2}%
\end{array}
& -\frac{\left( D-2\right) ^{1/2}}{2}\frac{1}{\kappa^{2}} \\ 
-\frac{\left( D-2\right) ^{1/2}}{2}\frac{1}{\kappa^{2}} & 0%
\end{array}
\right) ,  \label{non-degenerate matrices}
\end{equation}

\begin{equation}
b_{ij}\left( 0^{+}\right) =\left( 
\begin{array}{cc}
\frac{D}{2}\beta p^{2}-\frac{1}{\kappa^{2}}\left( 1-\frac{D}{2}\right)
+2\left( D-1\right) \alpha p^{2} & -i\sqrt{p^{2}}\frac{\left( D-2\right) }{%
\sqrt{2}\kappa^{2}} \\ 
i\sqrt{p^{2}}\frac{\left( D-2\right) }{\sqrt{2}\kappa^{2}} & 0%
\end{array}
\right) ,  \label{2}
\end{equation}

\begin{equation}
a\left( 2^{-}\right) =-\frac{1}{2\kappa^{2}}+2\gamma p^{2},
\end{equation}

\begin{equation}
a\left( 0^{-}\right) =\frac{1}{\kappa^{2}}+2\gamma p^{2};
\end{equation}

their respective inverses are listed in the sequel:

\begin{equation}
a_{ij}^{-1}\left( 2^{+}\right) =-\frac{2\kappa^{4}}{p^{2}}\left( 
\begin{array}{cc}
0 & -i\sqrt{p^{2}}\frac{1}{\sqrt{2}\kappa^{2}} \\ 
i\sqrt{p^{2}}\frac{1}{\sqrt{2}\kappa^{2}} & -\frac{1}{2\kappa^{2}}+\frac{1}{2%
}\beta p^{2}%
\end{array}
\right) ,  \label{2+ inverse matrix}
\end{equation}

\begin{equation}
b^{-1}\left( 1^{+}\right) =-2\kappa^{4}\left( 
\begin{array}{cc}
0 & \frac{1}{\sqrt{2}}\frac{1}{\kappa^{2}} \\ 
\frac{1}{\sqrt{2}}\frac{1}{\kappa^{2}} & \frac{1}{2\kappa^{2}}+\frac{1}{2}%
\beta p^{2}%
\end{array}
\right) ,  \label{1+ inverse matrix}
\end{equation}%
\begin{equation}
b_{ij}^{-1}\left( 1^{-}\right) =-\frac{4\kappa^{4}}{\left( D-2\right) }%
\left( 
\begin{array}{cc}
0 & \frac{\left( D-2\right) ^{1/2}}{2}\frac{1}{\kappa^{2}} \\ 
\frac{\left( D-2\right) ^{1/2}}{2}\frac{1}{\kappa^{2}} & \frac{\left(
D-2\right) }{2}\beta p^{2}~+~\frac{1}{\kappa^{2}}\left( \frac{D-3}{2}\right)
-2\left( D-3\right) \gamma p^{2}%
\end{array}
\right) ,  \label{1- inverse matrix}
\end{equation}

\begin{equation}
b_{ij}^{-1}\left( 0^{+}\right) =-\frac{2\kappa^{4}}{\left( D-2\right)
^{2}p^{2}}\left( 
\begin{array}{cc}
0 & i\sqrt{p^{2}}\frac{\left( D-2\right) }{\sqrt{2}\kappa^{2}} \\ 
-i\sqrt{p^{2}}\frac{\left( D-2\right) }{\sqrt{2}\kappa^{2}} & \frac{D}{2}%
\beta p^{2}-\frac{1}{\kappa^{2}}\left( 1-\frac{D}{2}\right) +2\left(
D-1\right) \alpha p^{2}%
\end{array}
\right) ,  \label{0+ inverse matrix}
\end{equation}

\begin{equation}
a^{-1}\left( 2^{-}\right) =\frac{1}{2\gamma\left( p^{2}-\frac{1}{%
4\gamma\kappa^{2}}\right) },  \label{2- inverse matrix}
\end{equation}

\begin{equation}
a^{-1}\left( 0^{-}\right) =\frac{1}{2\gamma\left( p^{2}+\frac{1}{%
2\gamma\kappa^{2}}\right) }.  \label{0- inverse matrix}
\end{equation}

We immediately get that there are two nonmassive poles in the 2$^{+}$, 0$%
^{+}$ sectors and two massive poles in the 2$^{-}$, 0$^{-}$ sectors.\ \
These results highlight a remarkable difference with respect to \cite{ja},
because we do not have spin-2 massive propagation for the vielbein; so, we
do not expect spin-2 massive graviton in any dimension. Actually, as it will be shown in the next section, if we impose unitarity, the model becomes trivial, in the sense that none of the modes can propagate. 
\newpage
\end{widetext}

\section{Toward\ a massive graviton}

Our initial motivation was to investigate the role of a propagating torsion
in the description of massive gravity. Ever since, our results are not
encouraging in the sense that, as seen from the previous analysis, there is
no room for the propagation of a massive graviton in our model.

From the inspection of the structure of the matrices (\ref{1}-\ref{2}), we
can understand how to cure this problem. In the curvature terms, we have
only contributions for $\omega\omega,\omega\phi,\phi\omega$ propagators.
Once the structure of the $\omega\phi,\phi\omega$ contributions are always
of the form $\sqrt{p^{2}}\times$(function of the constants $\kappa$, $\alpha$%
,$\ \beta$, and $D$), we do not expect that the determinant may exhibit
zeroes at$\ p^{2}=\mu^{2}\neq0$, which would correspond to massive poles.
So, we claim that the only way to get a massive pole is to insert a $%
\phi\phi $ contribution into these matrices. But, this is possible only if
we enlarge our initial Lagrangian with explicit torsion terms.

Within all possible quadratic terms that we can form with torsion, the
independent contributions turn out to be: $T_{\mu \nu }^{~~~~a}T_{~~~a}^{\mu
\nu }$, $T_{\mu a}^{~~~~a}T_{~~~b}^{\mu b}$, $T_{abc}T^{abc}$. For an
initial attempt, we could take a representative case and check that it does
the job we have in mind, namely, to introduce a massive pole. But, as we
wish to find possible unitary Lagrangian that describe massive gravitons, we
have aside our initial model and consider the most general
parity-preserving Lagrangian without higher derivatives, that is %
{\small
\begin{eqnarray}
\cal{L} & =&-\lambda R+\xi R^{2}+\left( s+t\right) R_{ab}R^{ab}+\left( s-t\right)
R_{ab}R^{ba} \nonumber\\
&  &+\frac{1}{6}\left( 2d+q\right) R_{abcd}R^{abcd}+\frac{1}{6}\left(
2d+q-6r\right) R_{abcd}R^{cdab}\nonumber\\
&  &+\frac{2}{3}\left(d-q\right) R_{abcd}R^{acbd}+\frac{1}{12}\left( 4u+v+3\lambda\right) T_{abc}T^{abc}\nonumber\\
&  &+\frac{1}{6}\left(-2u+v-3\lambda\right) T_{abc}T^{bca} \nonumber\\
&  &+\frac{1}{D-1}\left( -u+2w-\left( D-1\right)\lambda\right)
T_{ab}^{~~~~b}T_{~~~~c}^{ac}.  \label{complete lagrangian}
\end{eqnarray}}

The constant factors are chosen in this cumbersome way in order to simplify
the analysis of the conditions for unitarity. By linearizing $\cal{L}$
and, using the results of the Appendix, we write $\cal{L}$$_{2}$ in terms of the spin operators. The total linearized
Lagrangian can be written again as

\begin{equation}
\left( \cal{L}\right) _{2}=\displaystyle\sum_{\alpha,\beta,ij,J^{P}}%
\psi_{\alpha}a_{ij}^{\psi\lambda}\left( J^{P}\right) P_{ij}^{\psi\lambda
}\left( J^{P}\right) _{\alpha\beta}\lambda_{\beta}.
\end{equation}
But now, the coefficient matrices are given by

\begin{widetext}

\begin{equation}
a\left( 0^{+}\right) =\left( 
\begin{array}{ccc}
\left[ Ds+2d-2r+2\left( D-1\right) \xi\right]p^2+~w & -i%
\sqrt{2}\sqrt{p^2}w & 0 \\ 
i\sqrt{2}\sqrt{p^2}w & 2\left[ w-\left( \frac{D-2}{2}%
\lambda\right) \right]p^2 & 0 \\ 
0 & 0 & 0%
\end{array}
\right) ,
\end{equation}
{\small
\begin{gather}
{a\left( 1^{-}\right) = \left( 
\begin{array}{cccc}
\begin{array}{c}
\left[ \frac{\left(D-2\right) }{2}\left( s+t\right) +d\right]p^2 \\ 
+\frac{2\left( D-2\right) w+u}{2\left( D-1\right) }%
\end{array}
& -\frac{\left( D-2\right) ^{1/2}}{2}\left( \frac{2w-u}{D-1}\right) & -i%
\sqrt{p^{2}}\frac{\left( D-2\right) ^{1/2}}{2}\left( \frac{2w-u}{D-1}\right)
& -i\sqrt{p^{2}}\frac{\left( D-2\right) ^{1/2}}{2}\left( \frac{2w-u}{D-1}%
\right) \\ 
-\frac{\left( D-2\right) ^{1/2}}{2}\left( \frac{2w-u}{D-1}\right) & \frac{1}{%
D-1}\left[ w+u\left( \frac{D-2}{2}\right) \right] & i\sqrt{p^{2}}\frac{1}{D-1%
}\left[ w+u\left( \frac{D-2}{2}\right) \right] & i\sqrt {p^{2}}\frac{1}{D-1}%
\left[ w+u\left( \frac{D-2}{2}\right) \right] \\ 
i\sqrt{p^{2}}\frac{\left( D-2\right) ^{1/2}}{2}\left( \frac{2w-u}{D-1}\right)
& -i\sqrt{p^{2}}\frac{1}{D-1}\left[ w+u\left( \frac{D-2}{2}\right) \right] & 
p^{2}\frac{1}{D-1}\left[ w+u\left( \frac{D-2}{2}\right) \right] & p^{2}\frac{%
1}{D-1}\left[ w+u\left( \frac{D-2}{2}\right) \right] \\ 
i\sqrt{p^{2}}\frac{\left( D-2\right) ^{1/2}}{2}\left( \frac{2w-u}{D-1}\right)
& -i\sqrt{p^{2}}\frac{1}{D-1}\left[ w+u\left( \frac{D-2}{2}\right) \right] & 
p^{2}\frac{1}{D-1}\left[ w+u\left( \frac{D-2}{2}\right) \right] & p^{2}\frac{%
1}{D-1}\left[ w+u\left( \frac{D-2}{2}\right) \right]%
\end{array}
\right) , }
\end{gather}}

\begin{equation}
a\left( 2^{+}\right) =\left( 
\begin{array}{cc}
\left[ s+2d-2r\right]p^2+\frac{u}{2} & -i\sqrt{p^2}\frac{1}{\sqrt{2}}u \\ 
i\sqrt{p^2}\frac{1}{\sqrt{2}}u & \left( u+\lambda\right)p^2
\end{array}
\right)
\end{equation}

\begin{equation}
a\left( 2^{-}\right) =dp^2+\frac{u}{2},
\end{equation}

\begin{equation}
a\left( 1^{+}\right) =\left( 
\begin{array}{ccc}
\left( t-2r\right)p^2+\left( \frac{u+4v}{6}\right) & -\frac{%
\left( 2v-u\right) }{3\sqrt{2}} & -i\sqrt{p^{2}}\frac{\left( 2v-u\right) }{3%
\sqrt{2}} \\ 
-\frac{\left( 2v-u\right) }{3\sqrt{2}} & \frac{\left( u+v\right) }{3} & i%
\sqrt{p^{2}}\frac{\left( u+v\right) }{3} \\ 
i\sqrt{p^{2}}\frac{\left( 2v-u\right) }{3\sqrt{2}} & -i\sqrt{p^{2}}\frac{%
\left( u+v\right) }{3} & \frac{\left( u+v\right) }{3}p^{2}%
\end{array}
\right) ,
\end{equation}

\begin{equation}
a\left( 0^{-}\right) =qp^2+v.
\end{equation}

Again, we have degeneracies and, in order to obtain the saturated
propagator, we must pick out the nondegenerate submatrices. We quote their
inverses below:

\begin{equation}
a^{-1}\left( 2^{+}\right) =p^{-2}\left[ \left( s+2d-2r\right)
\left( u+\lambda\right)p^2+\frac{u}{2}\lambda\right]
^{-1}\left( 
\begin{array}{cc}
\left( u+\lambda\right)p^2 & i\sqrt{p^2}\frac{1%
}{\sqrt{2}}u \\ 
-i\sqrt{p^2}\frac{1}{\sqrt{2}}u & \left[ s+2d-2r\right] 
p^2+\frac{u}{2}%
\end{array}
\right) ,  \label{inverse coeficient matrix 2+}
\end{equation}

\begin{equation}
a^{-1}\left( 0^{-}\right) =\left( qp^2+v\right) ^{-1},
\end{equation}

\begin{equation}
a^{-1}\left( 2^{-}\right) =\left( dp^2+\frac{u}{2}\right)
^{-1},
\end{equation}

\begin{equation}
b^{-1}\left( 1^{+}\right) =\left[ \frac{1}{3}\left( t-2r\right) \left(
u+v\right)p^2+\frac{uv}{2}\right] ^{-1}\left( 
\begin{array}{cc}
\frac{\left( u+v\right) }{3} & \frac{\left( 2v-u\right) }{3\sqrt{2}} \\ 
\frac{\left( 2v-u\right) }{3\sqrt{2}} & \left( t-2r\right)p^2+\left(\frac{u+4v}{6}\right)%
\end{array}
\right) ,
\end{equation}

\begin{align}
b^{-1}\left( 1^{-}\right) & =\left\{ \frac{1}{D-1}\left[ \frac{\left(
D-2\right) }{2}\left( s+t\right) +p\right] \left[ w+u\left( \frac {D-2}{2}%
\right) \right]p^2+\frac{wu}{2}\right\} ^{-1}  \notag \\
& \times\left( 
\begin{array}{cc}
\frac{1}{D-1}\left[ w+u\left( \frac{D-2}{2}\right) \right] & \frac{\left(
D-2\right) ^{1/2}}{2}\left( \frac{2w-u}{D-1}\right) \\ 
\frac{\left( D-2\right) ^{1/2}}{2}\left( \frac{2w-u}{D-1}\right) & \left[ 
\frac{\left( D-2\right) }{2}\left( s+t\right) +p\right]p^2+
\frac{2\left( D-2\right) w+u}{2\left( D-1\right) }%
\end{array}
\right)
\end{align}

\begin{align}
b^{-1}\left( 0^{+}\right) & =p^{-2}\left[ \frac{1}{2}\left[
Ds+2d-2r+2\left( D-1\right) \xi\right] \left[ w-\frac{D-2}{2}\lambda%
\right]p^2-w\left( D-2\right) \lambda\right] ^{-1}  \notag \\
& \times\left( 
\begin{array}{cc}
2\left[ w-\frac{D-2}{2}\lambda\right]p^2 & i\sqrt{2}\sqrt{p^2}w \\ 
-i\sqrt{2}\sqrt{p^2}w & \left[ Ds+2d-2r+2\left( D-1\right)
\xi\right]p^2+w
\end{array}
\right)  \label{inverse coeficient matrix 0+}
\end{align}

We can now realize that our chances to describe a massive graviton have
enhanced. At the same time, the introduction of the new term endowed the
other spin sector with dynamics. So, apparently, we could obtain a
Pauli-Fierz analogue, that is, with only a spin-2 massive particle propagating, if we prevent the extra modes from propagating by imposing relations among the parameters of the Lagrangian, so that the extra poles vanish. But we do not investigate it. Our aim is only to give some insights on how to describe a
massive graviton whenever we have propagating torsion.
\end{widetext}

\section{Analysis of tachyonic and ghosts modes in the extended model}

Now, that we have obtained the inverses of the nondegenerate submatrices,
we can write the saturated propagator with an external current, $S_{\alpha}$:

\begin{equation}
\Pi=i\displaystyle\sum_{J^{P}} S_{\alpha}^{\ast}A_{ij}\left( J^{P}\right)
P_{ij}\left( J^{P}\right) _{\alpha\beta}S_{\beta}\left( p^{2}-m^{2}\right)
^{-1},  \label{satured propagator}
\end{equation}
where A$_{ij}$ are the matrices given above with the massive pole extracted.
So, these are $2\times2$ (or $1\times1$ for $1^{-}$ and $2^{-}$ sectors) matrices, which are degenerate at the pole.
According to Ref. \cite{bu}, for a massive
propagating particle not to be a tachyon or a ghost, we must require that%
\begin{equation}
m^{2}>0\ \ \ \ \text{and Im Res}\left( \Pi\right)
|_{p^{2}=m^{2}}>0,  \label{no tachyon condiction for massive case}
\end{equation}
which implies that, for each $J^{P}$,
we must have%
\begin{equation}
\left( -1\right) ^{P}trA\left( J^{P}\right) |_{p^{2}=m^{2}}>0.
\label{no ghost condiction for massive case}
\end{equation}
The $\left( -1\right) ^{p}$ factor comes from the evaluation of the spin
operators at the pole. The even (odd) operators have an even (odd) number of $%
\theta$ in their structure and each $\theta$\ contributes with a $\left(
-1\right) $ factor. For each spin we have the conditions:\newline\newline\newline
$\mathbf{2}^{+}:\ 2d-2r+s>0;\ u\lambda\left(u+\lambda\right) <0.$ 
\newline\newline
$\mathbf{2}^{-}:d<0;\ u>0.$ \newline\newline
$\mathbf{1}^{+}:2r+t>0;\ uv\left( u+v\right) <0.$ \newline\newline
$\mathbf{1}^{-}:\left[ \frac{\left( D-2\right) }{2}\left( s+t\right)+d
\right] <0;\ wu\left[ w+u\left( \frac{D-2}{2}\right) \right] >0.$ \newline\newline
$\mathbf{0}^{+}: \left[ Ds+2d-2r+2\left( D-1\right) \xi \right] >0;
\\ ~~~~~~~w\lambda\left[ w-\left( \frac{D-2}{2}\right)\lambda \right] >0.$\newline\newline
$\mathbf{0}^{-}:q<0;\ v>0.$
\newline\newline\newline

For the case of massless poles, the analysis requires extra care, because
there are new singularities, coming from the operators themselves, when we
evaluate them at the pole $p^{2}=0$. For this reason, we proceed in a
somewhat different form.

Because of the singularities of spin operators, even the matrices with massive poles
can contribute to the residue of the massless poles. The $p^{-6}$ and p$%
^{-4} $ singularities cancel out when we use the source constraints. It can
be shown that from all $p^{-2}$ singularities, only those associated with
the Einstein-Hilbert survives. The final result to the residue of massless
poles is
\begin{eqnarray}
\text{Im Res}\left( \Pi \right) _{p^{2}=0} &=&\frac{\lambda ^{-1}}{%
p^{2}}\left( 
\begin{array}{cc}
\tau ^{ab\ast } & \Sigma ^{ab\ast }%
\end{array}%
\right) \left( 
\begin{array}{cc}
4 & 2i\notag \\ 
-2i & 1%
\end{array}%
\right)\notag \\
&&\times \left[ P\left( 2^{+},\eta \right) -\frac{1}{D-2}P\left( 0^{+},\eta
\right) \right]\notag \\
&&\times \left( 
\begin{array}{c}
\tau ^{cd}\label{massless propagator for complete lagrangian} \\ 
\Sigma ^{cd}%
\end{array}%
\right). 
\end{eqnarray}

As the matrix that appears in this equation is Hermitean, it can be diagonalized by a suitable unitary matrix. Making this change of variables, we can rewrite this expression as

\begin{equation}
\text{Im Res}\left( \Pi \right) _{p^{2}=0} =5\frac{\lambda ^{-1}}{%
p^{2}}\left(\tilde{\tau} ^{(ab)\ast}\tilde{\tau} _{(ab)}-\frac{1}{D-2}\tilde{\tau} ^{a\ast}_a\tilde{\tau} ^{b}_b
\right). 
\end{equation}.

Furthermore, choosing a suitable basis in the D-dimensional Minkoswski space, we can expand the source $\tau^{(ab)}$ as 

\begin{align}
\tau ^{(ab)} &=c_1p^ap^b+c_{2\alpha}(p^b\epsilon ^{a\alpha}+p^b\epsilon^{a\alpha})\notag \\
&+c_{3\alpha\beta}(\epsilon ^{a\alpha}\epsilon ^{b\beta}) \label{tau source in terms of momenta}
\end{align}
where,

\begin{subequations}
\begin{align}
p^a & =(p_0,\vec p), \\
q^a & =(p_0,-\vec p), \\
\epsilon_{\alpha}^a &,\quad \alpha=1,...,D-2
\end{align}
with,
\begin{align}
p^2 &=q^2=0, \\
p.q &=(p_0)^2+(\vec p)^2\neq0, \\
p.\epsilon_{\alpha} &=q.\epsilon_{\alpha}=0, \\
\epsilon_{\alpha}.\epsilon_{\beta} &=-\delta_{\alpha\beta}.
\end{align}
\end{subequations}
These vectors span the D-dimensional Minkowski space. Plugging (\ref{tau source in terms of momenta}) in (\ref{massless propagator for complete lagrangian}), we obtain

\begin{equation}
\text{Im Res}\left( \Pi \right) _{p^{2}=0} =5\frac{\lambda ^{-1}}{%
p^{2}}\left(c^{\ast}_{3\alpha\beta}c^{\alpha\beta}_3-\frac{1}{D-2}c^{\alpha\ast}_{3\alpha}c^{\beta}_{3\beta}
\right). 
\end{equation}.

Let us relabel the ($D-2$) ${c_3}^{\prime}$s by $c_i$, with $i=1,...,D-2=N$. So, this expression can be written as:
\begin{equation}
\text{Im Res}\left( \Pi \right) _{p^{2}=0} =5\frac{\lambda ^{-1}}{%
p^{2}}\displaystyle\sum_{i>j=1}^{N}\frac{1}{N}\left(|c_i|^2+|c_j|^2-(c^{\ast}_ic_j+c^{\ast}_jc_i)
\right). 
\end{equation}.

This expression vanishes for $D=3$ $(N=1)$ and is positive-definite for $D>3$ $(N>1)$ if the condition $\lambda=\frac{1}{\kappa^{2}}>0$ is chosen. As the above matrix is degenerate, there is only one propagating mode.

We are now in a position to analyze the spectrum of the initial model of Sec. II. Comparing the two Lagrangians, (\ref{lagrangian}) and (\ref{complete lagrangian}), we see that both agree if we identify

\begin{eqnarray}
\lambda &=& \frac{1}{\kappa^2};\  \  \  \ \beta-4\gamma=s+t;\nonumber \\
\alpha+\gamma &=& \xi;\  \  \  \ \gamma=\frac{d}{2}=\frac{q}{2}=r \nonumber \\
u &=& -v=-\lambda;\  \  \  \  \frac{D-2}{2}u+w=0.
\end{eqnarray}

We see, from the matrices (\ref{inverse coeficient matrix 2+})-(\ref{inverse coeficient matrix 0+}), that these relations confirm our previous result that there are only massive poles in the $2^-$ and $0^-$ sectors. Furthermore, the conditions in the parameters for the full Lagrangian, derived above, tell us that the propagation of these modes are incompatible, since $d=q$. Therefore, the initial model has no propagation mode for $D=3$ and has only a massless graviton propagating for $D>3$.

Now, we must search for the possible unitary Lagrangians resulting from the
possible intersections of the conditions above for the extended Lagrangian. However, there is a net
conflict among these relations. Namely, the conditions for the $2^{-}$ sector requires $u>0$, whereas the conditions for $2^{+}$ impose $u<0$.
Therefore, for arbitrary values of the parameters in the Lagrangian (\ref%
{complete lagrangian}), we have a nonunitary model. In order to achieve
unitarity for a propagating massive graviton, we must assume that some modes
do not propagate. In so doing, the conditions related to these modes do not
need to be satisfied. The conditions for the nonpropagation of a mode are
readily seen from the matrices (\ref{inverse coeficient matrix 2+})-(\ref%
{inverse coeficient matrix 0+}). They are obtained by requiring the absence
of a pole related to the mode. In the sequel, we present the conditions that
must be satisfied in order to get a unitary model with a propagating massive
graviton:

\begin{enumerate}
\item[i)] $d=0;\ \ 2r+t<0; \ s=-t;\ u=-v;\\ \frac{1}{\kappa^{2}}>v>0;\ q<0;\ w\left( w-\frac{D-2}{2\kappa ^{2}}\right) >0%
\\ -Dt-2r+2\left( D-1\right) \xi >0.$
\end{enumerate}

This corresponds to the following Lagrangian:%
\begin{align}
\cal{L} & =-\frac{1}{\kappa ^{2}}R+\xi R^{2}-2tR_{ab}R^{ba}+\frac{1}{6}%
qR_{abcd}R^{abcd}  \notag \\
& +\frac{1}{6}\left( q-6r\right) R_{abcd}R^{cdab}-\frac{2}{3}%
qR_{abcd}R^{acbd}  \notag \\
& +\frac{1}{4}\left( -v+\frac{1}{\kappa ^{2}}\right) \left(
T_{abc}T^{abc}-2T_{abc}T^{bca}\right)   \notag \\
& +\frac{1}{D-1}\left( v+2w-\frac{\left( D-1\right) }{\kappa ^{2}}\right)
T_{ab}^{~~~~b}T_{~~~~c}^{ac},  \label{first unitary lagrangian}
\end{align}%
where the parameters satisfy the conditions of item i. The propagating modes
are: a spin-$2^{+}$ massless (for $D\geq 4$), and massives spin-$2^{+}$,
spin-$0^{+}$, spin-$0^{-}$. There are several particular cases of (\ref%
{first unitary lagrangian}) corresponding to inhibition of the propagation
of the modes $0^{+}$ and $0^{-}$.

\begin{enumerate}
\item[ii)] $d=0;\ -2r+s>0;\ \ t=-2r;\\
w=-u\left( \frac{D-2}{2}\right);\ \ -\frac{1}{\kappa ^{2}}<u<0;\\
Ds-2r+2\left( D-1\right) \xi =0;q<0;v>0.$
\end{enumerate}

With these parameters, we have the second unitary Lagrangian:%
\begin{align}
\cal{L} & =-\frac{1}{\kappa ^{2}}R+\frac{2r-Ds}{2\left( D-1\right) }R^{2}+\left(
s-2r\right) R_{ab}R^{ab}  \notag \\
& +\left( s+2r\right) R_{ab}R^{ba}+\frac{1}{6}qR_{abcd}R^{abcd}  \notag \\
& +\frac{1}{6}\left( q-6r\right) R_{abcd}R^{cdab}-\frac{2}{3}%
qR_{abcd}R^{acbd}  \notag \\
& +\frac{1}{12}\left( 4u+v+\frac{3}{\kappa ^{2}}\right) T_{abc}T^{abc} 
\notag \\
& +\frac{1}{6}\left( -2u+v-\frac{3}{\kappa ^{2}}\right) T_{abc}T^{bca} 
\notag \\
& -\left( u+\frac{1}{\kappa ^{2}}\right) T_{ab}^{~~~~b}T_{~~~~c}^{ac}.
\end{align}%
In addition to the massive graviton, this model carries the massless
graviton (for $D\geq 4$) and a $0^{-}$ massive particle.

\begin{enumerate}
\item[iii)] $d=0;\ -2r+s>0;\ u=-v;\ s+t<0;\\ 2r+t<0;\ \ w\left(w-v\left( \frac{D-2}{2}\right)\right)<0 ;\ \frac{1}{\kappa ^{2}}>v>0;\\
q<0;\ \ Ds-2r+2\left( D-1\right) \xi =0.$
\end{enumerate}

The related Lagrangian is

\begin{align}
\cal{L} & =-\lambda R+\frac{2r-Ds}{2\left( D-1\right) }R^{2}+\left( s+t\right)
R_{ab}R^{ab}  \notag \\
& \frac{1}{6}qR_{abcd}R^{abcd}+\frac{1}{2}\left( v-\frac{1}{\kappa ^{2}}%
\right) T_{abc}T^{bca}  \notag \\
& +\frac{1}{6}\left( q-6r\right) R_{abcd}R^{cdab}-\frac{2}{3}%
qR_{abcd}R^{acbd}  \notag \\
& +\frac{1}{4}\left( -v+\frac{1}{\kappa ^{2}}\right) T_{abc}T^{abc}+\left(
s-t\right) R_{ab}R^{ba}  \notag \\
& +\frac{1}{D-1}\left( v+2w-\frac{\left( D-1\right) }{\kappa ^{2}}\right)
T_{ab}^{~~~~b}T_{~~~~c}^{ac}.
\end{align}%
This model propagates the massless (for $D\geq 4$) and the massive graviton,
along with massive $1^{-}$ and $0^{-}$ particles.

\begin{enumerate}
\item[iv)] $d=0;\ 2r+t>0;\ -2r+s>0;\ w=-u\left( \frac{D-2}{2}\right);\\
-\frac{1}{\kappa ^{2}}<u<0;\ \xi =\frac{2r-Ds}{2\left( D-1\right) };\ q<0;\ v>0;\ \ u>-v.$
\end{enumerate}

These conditions exhaust our possibilities of describing a massive graviton
in a unitary way. The Lagrangian associated is given by

\begin{align}
\cal{L} & =-\lambda R+\frac{2r-Ds}{2\left( D-1\right) }R^{2}+\left( s+t\right)
R_{ab}R^{ab}  \notag \\
& -\left( u-\frac{1}{\kappa ^{2}}\right) T_{ab}^{~~~~b}T_{~~~~c}^{ac}.+\frac{%
1}{6}qR_{abcd}R^{abcd}  \notag \\
& +\frac{1}{6}\left( q-6r\right) R_{abcd}R^{cdab}-\frac{2}{3}%
qR_{abcd}R^{acbd}  \notag \\
& +\frac{1}{12}\left( 4u+v+\frac{3}{\kappa ^{2}}\right) T_{abc}T^{abc} 
\notag \\
& +\frac{1}{6}\left( -2u+v-\frac{3}{\kappa ^{2}}\right)
T_{abc}T^{bca}+\left( s-t\right) R_{ab}R^{ba}
\end{align}%
In addition to the massive and massless graviton (for $D\geq 4$), there are
massive $1^{+}$ and $0^{-}$ dynamical particles.
As we have pointed out in the Introduction, there is considerable interest in $D=3$ gravity. So, it is worthy to stress that our results are valid for this dimension too. With the exception that in this dimension there is no propagating massless spin-$2^{+}$, the other analyzed features are essentially the same.
In order to compare with the work of Ref. \cite{bu}, we investigate these
conditions for $D=4$. In this case, we have the following unitary
Lagrangians:

\begin{widetext}

\begin{enumerate}
\item[i)] $\ $$\cal{L}$$ =-\lambda R-2tR_{ab}R^{ba}+\frac{1}{6}%
qR_{abcd}R^{abcd}+\frac{1}{6}\left( q-6r\right) R_{abcd}R^{cdab}-\frac{2}{3}%
qR_{abcd}R^{acbd}\newline
\ \ \ \ +\frac{1}{4}\left( -v+\frac{1}{\kappa ^{2}}\right) \left(
T_{abc}T^{abc}-2T_{abc}T^{bca}\right) +\frac{1}{3}\left( v+2w-\frac{3}{%
\kappa ^{2}}\right) T_{ab}^{~~~~b}T_{~~~~c}^{ac},$
\end{enumerate}

with $r<-\frac{t}{2};~\frac{1}{\kappa^{2}}>v>0;~q<0;~r<-2t;~w\left( w-\frac{1%
}{\kappa^{2}}\right) >0.$

\begin{enumerate}
\item[ii)] $\ $$\cal{L}$$=-\lambda R-3sR_{ab}R^{ab}+5sR_{ab}R^{ba}+\frac{1}{6%
}qR_{abcd}R^{abcd}+\frac{1}{6}\left( q-12s\right) R_{abcd}R^{cdab}-\frac{2}{3%
}qR_{abcd}R^{acbd}\newline
+\frac{1}{12}\left( 4u+v+\frac{3}{\kappa^{2}}\right) T_{abc}T^{abc}+\frac{1}{%
6}\left( -2u+v-\frac{3}{\kappa^{2}}\right) T_{abc}T^{bca}-\left( u+\frac{1}{%
\kappa^{2}}\right) T_{ab}^{~~~~b}T_{~~~~c}^{ac},$
\end{enumerate}

with $-\frac{1}{\kappa^{2}}<u<0;~q<0;~s<0;~v>0.$

\begin{enumerate}
\item[iii)] $\ $$\cal{L}$$ =-\lambda R+\left( s+t\right) R_{ab}R^{ab}+\left(
s-t\right) R_{ab}R^{ba}+\frac{1}{6}qR_{abcd}R^{abcd}+\frac{1}{6}\left(
q-12s\right) R_{abcd}R^{cdab}\newline
-\frac{2}{3}qR_{abcd}R^{acbd}+\frac{1}{4}\left( -v+\frac{1}{\kappa ^{2}}%
\right) T_{abc}T^{abc}+\frac{1}{2}\left( v-\frac{1}{\kappa ^{2}}\right)
T_{abc}T^{bca}+\frac{1}{3}\left( v+2w-\frac{3}{\kappa ^{2}}\right)
T_{ab}^{~~~~b}T_{~~~~c}^{ac},$
\end{enumerate}

with $s<0;~s+t<0;~w(w-v)<0;~\frac{1}{\kappa^{2}}>v>0;~q<0.$

\begin{enumerate}
\item[iv)] $\ $$\cal{L}$$ =-\lambda R+\left( s+t\right) R_{ab}R^{ab}+\left(
s-t\right) R_{ab}R^{ba}+\frac{1}{6}qR_{abcd}R^{abcd}+\frac{1}{6}\left(
q-6r\right) R_{abcd}R^{cdab}\newline
-\frac{2}{3}qR_{abcd}R^{acbd}+\frac{1}{12}\left( 4u+v+\frac{3}{\kappa ^{2}}%
\right) T_{abc}T^{abc}+\frac{1}{6}\left( -2u+v-\frac{3}{\kappa ^{2}}\right)
T_{abc}T^{bca}-\left( u-\frac{1}{\kappa ^{2}}\right)
T_{ab}^{~~~~b}T_{~~~~c}^{ac},$
\end{enumerate}

with $2r+t>0;~s<0;~-\frac{1}{\kappa^{2}}<u<0;~q<0;~v>0.$

\bigskip To get these results, we have taken $\xi=0$ due to the
four-dimensional version of the Gauss-Bonet theorem, which states that for
asymptotically flat spaces:

\begin{equation}
\int d^{4}xe\left( R_{\mu \nu ab}R^{\mu \nu ab}-4R_{\mu a}R^{\mu
a}+R^{2}\right) =0  \label{gauss-bonet}
\end{equation}%
and so, there is no need for the presence of the three corresponding terms in
above Lagrangians. These conditions correspond to particular cases of those
listed in \cite{bu}. Therefore, all Lagrangians that describe a massive
graviton in a unitary way in D dimensions are reduced to some particular
case already mentioned in \cite{bu}.

\end{widetext}

\section{Concluding Remarks}

We set our discussion by trying to generalize the results of \cite{ja} for
the case of a nontrivial and propagating torsion. We conclude that our
naive ansatz of simply considering the same form of that Lagrangian is not
sufficient to describe massive gravitons, and the requirement of unitarity
is so severe that the model becomes trivial. Some considerations guide us to
the conclusion that, if we wish to introduce massive gravitons, we should
include explicit torsion terms in the Lagrangian. Furthermore, as we are
interested in the analysis of unitarity, we consider the most general
parity-preserving Lagrangian in D dimensions without higher derivatives, and
we investigate the constraints on the parameters so as to ensure the
unitarity. We find a set of unitary Lagrangians in D dimensions that
propagate a massive graviton, and we verify that these Lagrangians agree
with those listed in \cite{bu}, in the particular case of $D=4$. But, for $%
D\neq 4$, as the Gauss-Bonet theorem includes products of more than two
curvature-type tensors, there are many more conditions compatible with the
unitarity constraints on the parameters, due to the extra parameter $\alpha $%
.

The initial purpose was partly reached, once we have found unitary
Lagrangians with propagating torsion that describes at least a massive
graviton. However, what we have done is not quite a generalization of the
results of \cite{ja}, since there, the linearized Lagrangian corresponds to
the Pauli-Fierz Lagrangian, which is intrinsically defined in the second-order formalism for gravitation. We could try to define such a model in case
the torsion propagates by inhibiting the propagation of all the other
modes, but the massive graviton. However, we are here more interested on the
considerations that should be made to shed some light on models for massive
gravitons, whenever we consider a more fundamental approach to gravitation
(in the sense of gauge theories). The lesson we draw is that torsion
actually plays a crucial role in the discussion, confirming previous results
we have referred to in the course of this paper.

\textbf{ACKNOWLEDGMENTS:}

We are grateful to R. Nardi and M. V. S. Fonseca for constructive
discussions. The authors express their gratitude to CNPq-Brazil and FAPERJ
for their invaluable financial support.

\begin{widetext}

\textbf{APPENDIX: SPIN PROJECTORS}

\begin{enumerate}
\item $P_{21}^{\phi\omega}\left(0^{+}\right) _{ab;def}=\frac{\sqrt{2}}{%
2\left( D-1\right) }\left(
\theta_{ab}\theta_{de}p_{f}-\theta_{ab}\theta_{df}p_{e}\right) $

\item $P_{31}^{\phi\omega}\left( 0^{+}\right) _{ab;def}=\frac{\sqrt{2}}{%
2\left( D-1\right) ^{1/2}}\left( \omega_{ab}\theta_{de}p_{f}-\omega
_{ab}\theta_{df}p_{e}\right) $

\item $P_{12}^{\omega\phi}\left( 0^{+}\right) _{abc;df}=\frac{\sqrt{2}}{%
2\left( D-1\right) }\left(
\theta_{ab}\theta_{df}p_{c}-\theta_{ac}\theta_{df}p_{b}\right) $

\item $P_{13}^{\omega\phi}\left( 0^{+}\right) _{abc;df}=\frac{\sqrt{2}}{%
2\left( D-1\right) ^{1/2}}\left( \theta_{ab}p_{c}-\theta_{ac}p_{b}\right)
\omega_{df}~$

\item $P_{22}^{\phi\phi}\left( 0^{+}\right) {}_{ab;cd}=\frac{1}{D-1}%
\theta_{ab}\theta_{cd}$

\item $P_{33}^{\phi\phi}\left( 0^{+}\right)
{}_{ab;cd}=\omega_{ab}\omega_{cd} $

\item $P_{23}^{\phi\phi}\left( 0^{+}\right) _{ab;cd}=\frac{1}{\sqrt{D-1}}%
\theta_{ab}\omega_{cd}$

\item $P_{32}^{\phi\phi}\left( 0^{+}\right) _{ab;cd}=\frac{1}{\sqrt{D-1}}%
\omega_{ab}\theta_{cd}$

\item $P_{11}^{\omega\omega}\left( 0^{+}\right) _{abc;def}=\frac{1}{2\left(
D-1\right) }\left[ \theta_{ab}\left( \theta_{de}\omega_{cf}-\theta
_{df}\omega_{ce}\right) -\theta_{ac}\left(
\theta_{de}\omega_{bf}-\theta_{df}\omega_{be}\right) \right] $

\item $P_{11}^{\omega\omega}(1^{-})~_{abc;def}=\frac{1}{2\left( D-2\right) }%
\left[ \theta_{ab}\left( \theta_{de}\theta_{cf}-\theta_{df}\theta
_{ce}\right) -\theta_{ac}\left(
\theta_{de}\theta_{bf}-\theta_{df}\theta_{be}\right) \right] $

\item $P_{31}^{\phi\omega}\left( 1^{-}\right) _{ab;def}=-\frac{1}{2\left(
D-2\right) ^{1/2}}\left( \theta_{de}\theta_{af}p_{b}+\theta_{de}\theta
_{bf}p_{a}-\theta_{df}\theta_{ae}p_{b}-\theta_{df}\theta_{be}p_{a}\right) $

\item $P_{32}^{\phi\omega}\left( 1^{-}\right) _{ab;def}=\frac{1}{2}\left[
\omega_{ad}\left( \theta_{bf}p_{e}-\theta_{be}p_{f}\right) +\omega
_{bd}\left( \theta_{af}p_{e}-\theta_{ae}p_{f}\right) \right] $

\item $P_{13}^{\omega\phi}\left( 1^{-}\right) _{abc;df}=\frac{1}{2\left(
D-2\right) ^{1/2}}\left( \theta_{ac}\theta_{bd}p_{f}+\theta_{ac}\theta
_{bf}p_{d}-\theta_{ab}\theta_{cd}p_{f}-\theta_{ab}\theta_{cf}p_{d}\right) $

\item $P_{23}^{\omega\phi}\left( 1^{-}\right) _{abc;df}=\frac{1}{2}\left[
\omega_{da}\left( \theta_{fc}p_{b}-\theta_{fb}p_{c}\right) +\omega
_{fa}\left( \theta_{dc}p_{b}-\theta_{db}p_{c}\right) \right] $

\item $P_{41}^{\chi\omega}\left( 1^{-}\right) _{ab;def}=\frac{1}{2\left(
D-2\right) ^{1/2}}\left( \theta_{ae}\theta_{df}p_{b}-\theta_{be}\theta
_{df}p_{a}-\theta_{af}\theta_{de}p_{b}+\theta_{bf}\theta_{de}p_{a}\right) $

\item $P_{42}^{\chi\omega}\left( 1^{-}\right) _{ab;def}=\frac{1}{2}\left[
\omega_{ad}\left( \theta_{be}p_{f}-\theta_{bf}p_{e}\right) +\omega
_{bd}\left( \theta_{af}p_{e}-\theta_{ae}p_{f}\right) \right] $

\item $P_{14}^{\omega\chi}\left( 1^{-}\right) _{abc;df}=\frac{1}{2\left(
D-2\right) ^{1/2}}\left( \theta_{ac}\theta_{bd}p_{f}-\theta_{ab}\theta
_{cd}p_{f}-\theta_{ac}\theta_{bf}p_{d}+\theta_{ab}\theta_{cf}p_{d}\right) $

\item $P_{24}^{\omega\chi}\left( 1^{-}\right) _{abc;df}=\frac{1}{2}\left[
\omega_{da}\left( \theta_{fb}p_{c}-\theta_{fc}p_{b}\right) +\omega
_{fa}\left( \theta_{dc}p_{b}-\theta_{db}p_{c}\right) \right] $

\item $P_{33}^{\phi\phi}\left( 1^{-}\right) {}_{ab;cd}=\frac{1}{2}\left(
\theta_{ac}\omega_{bd}+\theta_{bc}\omega_{ad}+\theta_{ad}\omega_{bc}+%
\theta_{bd}\omega_{ac}\right) $

\item $P_{44}^{\chi\chi}\left( 1^{-}\right) {}{}{}_{ab;cd}=\frac{1}{2}\left(
\theta_{ac}\omega_{bd}-\theta_{ad}\omega_{bc}-\theta_{bc}\omega_{ad}+%
\theta_{bd}\omega_{ac}\right) $

\item $P_{34}^{\phi\chi}\left( 1^{-}\right) {}_{ab;cd}=\frac{1}{2}\left(
\theta_{ac}\omega_{bd}-\theta_{ad}\omega_{bc}+\theta_{bc}\omega_{ad}-%
\theta_{bd}\omega_{ac}\right) $

\item $P_{43}^{\chi\phi}\left( 1^{-}\right) {}_{ab;cd}=\frac{1}{2}\left(
\theta_{ac}\omega_{bd}+\theta_{ad}\omega_{bc}-\theta_{bc}\omega_{ad}-%
\theta_{bd}\omega_{ac}\right) $

\item $P_{22}^{\omega\omega}(1^{-})_{abc;def}=\frac{1}{2}\omega_{ad}\left[
\theta_{be}\omega_{cf}-\theta_{bf}\omega_{ce}-\theta_{ce}\omega_{bf}+%
\theta_{cf}\omega_{be}\right] $

\item $P_{12}^{\omega\omega}\left( 1^{-}\right) _{abc;def}=\frac{1}{2\left(
D-2\right) ^{1/2}}\left\{ \theta_{ab}\left[ \theta_{ce}\omega_{df}-%
\theta_{cf}\omega_{de}\right] -\theta_{ac}\left[ \theta_{be}\omega
_{df}-\theta_{bf}\omega_{de}\right] \right\} $

\item $P_{21}^{\omega\omega}\left( 1^{-}\right) _{abc;def}=\frac{1}{2\left(
D-2\right) ^{1/2}}\left\{ \omega_{ab}\left[ \theta_{df}\theta_{ce}-%
\theta_{de}\theta_{cf}\right] -\omega_{ac}\left[ \theta_{df}\theta
_{be}-\theta_{de}\theta_{bf}\right] \right\} $

\item $P_{21}^{\phi \omega }\left( 2^{+}\right) _{ab;def}=\frac{\sqrt{2}}{4}%
\left[ \left( \theta _{ad}\theta _{be}+\theta _{ae}\theta _{bd}-\frac{2}{%
\left( D-1\right) }\theta _{ab}\theta _{de}\right) p_{f}-\left( \theta
_{ad}\theta _{bf}+\theta _{af}\theta _{bd}-\frac{2}{\left( D-1\right) }%
\theta _{ab}\theta _{df}\right) p_{e}\right] $

\item $P_{12}^{\omega \phi }\left( 2^{+}\right) _{abc;df}=\frac{\sqrt{2}}{4}%
\left[ \left( \theta _{ad}\theta _{bf}+\theta _{af}\theta _{bd}-\frac{2}{%
\left( D-1\right) }\theta _{ab}\theta _{df}\right) p_{c}-\left( \theta
_{ad}\theta _{cf}+\theta _{af}\theta _{cd}-\frac{2}{\left( D-1\right) }%
\theta _{ac}\theta _{df}\right) p_{b}\right] $

\item $P_{22}^{\phi \phi }\left( 2^{+}\right) {}_{ab;cd}=\frac{1}{2}\left(
\theta _{ac}\theta _{bd}+\theta _{ad}\theta _{bc}\right) -\frac{1}{\left(
D-1\right) }\theta _{ab}\theta _{cd}$

\item $P_{11}^{\omega \omega }\left( 2^{+}\right) _{abc;def}=\frac{1}{2}%
\theta _{ad}\left( \theta _{be}\omega _{cf}-\theta _{bf}\omega _{ce}-\theta
_{ce}\omega _{bf}+\theta _{cf}\omega _{be}\right) -\frac{1}{4}\theta _{ad}%
\left[ \theta _{be}\omega _{cf}-\theta _{bf}\omega _{ce}-\theta _{ce}\omega
_{bf}+\theta _{cf}\omega _{be}\right] $

$+\frac{1}{4}\theta _{ae}\left[ \theta _{cd}\omega _{bf}-\theta _{bd}\omega
_{cf}\right] +\frac{1}{4}\theta _{af}\left[ \theta _{bd}\omega _{ce}-\theta
_{cd}\omega _{be}\right] -\frac{1}{2\left( D-1\right) }\left[ \theta
_{ab}\left( \theta _{de}\omega _{cf}-\theta _{df}\omega _{ce}\right) -\theta
_{ac}\left( \theta _{de}\omega _{bf}-\theta _{df}\omega _{be}\right) \right] 
$

\item $P_{31}^{\chi \omega }\left( 1^{+}\right) _{ab;def}=\frac{\sqrt{2}}{4}%
\left( \theta _{af}\theta _{bd}k_{e}-\theta _{bd}\theta _{ae}p_{f}+\theta
_{ad}\theta _{be}p_{f}-\theta _{ad}\theta _{bf}p_{e}\right) $

\item $P_{32}^{\chi \omega }\left( 1^{+}\right) _{ab;def}=-\frac{1}{2}\left(
\theta _{bf}\theta _{ae}-\theta _{af}\theta _{be}\right) p_{d}~~~~$

\item $P_{13}^{\omega \chi }\left( 1^{+}\right) _{abc;df}=\frac{\sqrt{2}}{4}%
\left( \theta _{af}\theta _{cd}p_{b}-\theta _{bd}\theta _{af}p_{c}+\theta
_{ad}\theta _{bf}p_{c}-\theta _{ad}\theta _{cf}p_{b}\right) $

\item $P_{23}^{\omega \chi }\left( 1^{+}\right) _{abc;df}=-\frac{1}{2}\left(
\theta _{bd}\theta _{cf}-\theta _{cd}\theta _{bf}\right) p_{a}$

\item $P_{33}^{\chi \chi }\left( 1^{+}\right) {}{}_{ab;cd}=\frac{1}{2}\left(
\theta _{ac}\theta _{bd}-\theta _{ad}\theta _{bc}\right) $

\item $P_{22}^{\omega \omega }(1^{+})~_{abc;def}=\frac{1}{2}\omega _{ad}%
\left[ \theta _{be}\theta _{cf}-\theta _{bf}\theta _{ce}\right] $

\item $P_{12}^{\omega \omega }\left( 1^{+}\right) _{abc;def}=-\frac{\sqrt{2}%
}{4}\left[ \theta _{ae}\left( \theta _{bf}\omega _{cd}-\theta _{cf}\omega
_{bd}\right) -\theta _{af}\left( \theta _{be}\omega _{cd}-\theta _{ce}\omega
_{bd}\right) \right] $

\item $P_{21}^{\omega \omega }\left( 1^{+}\right) _{abc;def}=-\frac{\sqrt{2}%
}{4}\left[ \theta _{cd}\left( \theta _{bf}\omega _{ae}-\theta _{be}\omega
_{af}\right) +\theta _{bd}\left( \theta _{ce}\omega _{af}-\theta _{cf}\omega
_{ae}\right) \right] $

\item $P_{11}^{\omega \omega }(1^{+})_{abc;def}=\frac{1}{4}\theta _{ad}\left[
\theta _{be}\omega _{cf}-\theta _{bf}\omega _{ce}-\theta _{ce}\omega
_{bf}+\theta _{cf}\omega _{be}\right] +\frac{1}{4}\theta _{ae}\left[ \theta
_{cd}\omega _{bf}-\theta _{bd}\omega _{cf}\right] +\frac{1}{4}\theta _{af}%
\left[ \theta _{bd}\omega _{ce}-\theta _{cd}\omega _{be}\right] $

\item $P^{\omega \omega }\left( 0^{-}\right) _{abc;def}=\frac{1}{6}\left\{
\theta _{ad}\left[ \theta _{be}\theta _{cf}-\theta _{bf}\theta _{ce}\right]
+\theta _{ae}\left[ \theta _{bf}\theta _{cd}-\theta _{bd}\theta _{cf}\right]
+\theta _{af}\left[ \theta _{bd}\theta _{ce}-\theta _{be}\theta _{cd}\right]
\right\} $

\item $P^{\omega \omega }\left( 2^{-}\right) _{abc;def}=\frac{1}{2}\theta
_{ad}\left( \theta _{be}\theta _{cf}-\theta _{bf}\theta _{ce}\right) -\frac{1%
}{2\left( D-2\right) }\left[ \theta _{ab}\left( \theta _{de}\theta
_{cf}-\theta _{df}\theta _{ce}\right) -\theta _{ac}\left( \theta _{de}\theta
_{bf}-\theta _{df}\theta _{be}\right) \right]\\
\ ~~~~~~~~~~~~~~~~~~~~~~~~ -\frac{1}{6}\left\{
\theta _{ad}\left[ \theta _{be}\theta _{cf}-\theta _{bf}\theta _{ce}\right]
+\theta _{ae}\left[ \theta _{bf}\theta _{cd}-\theta _{bd}\theta _{cf}\right]
+\theta _{af}\left[ \theta _{bd}\theta _{ce}-\theta _{be}\theta _{cd}\right]
\right\} $
\end{enumerate}

\bigskip \end{widetext}


\begin{thebibliography}{99}

\bibitem{lhc1}
  K.~Hagiwara, P.~Konar, Q.~Li, K.~Mawatari and D.~Zeppenfeld,
  %``Graviton production with 2 jets at the LHC in large extra dimensions,''
  JHEP {\bf 0804} (2008) 019
  [arXiv:0801.1794 [hep-ph]].
  %%CITATION = JHEPA,0804,019;%%

\bibitem{lhc2}
  Z.~Ya-Jin, M.~Wen-Gan, H.~Liang and Z.~Ren-You,
  %``Associated production of graviton with $e^+e^-$ pair via photon-photon
  %collisions at a linear collider,''
  Phys.\ Rev.\  D {\bf 76} (2007) 054003
  [arXiv:0708.1195 [hep-ph]].
  %%CITATION = PHRVA,D76,054003;%%

\bibitem{lhc3}
  C.~Cartier, R.~Durrer and M.~Ruser,
  %``On graviton production in braneworld cosmology,''
  Phys.\ Rev.\  D {\bf 72} (2005) 104018
  [arXiv:hep-th/0510155].
  %%CITATION = PHRVA,D72,104018;%%

\bibitem{lhc4}
  P.~Jain and S.~Panda,
  %``Two graviton production at e+ e- and hadron hadron colliders in the
  %Randall-Sundrum model,''
  JHEP {\bf 0403} (2004) 011
  [arXiv:hep-ph/0401222].
  %%CITATION = JHEPA,0403,011;%%


\bibitem{nieto} A. S. Goldhaber, M. M. Nieto, "Photon and Graviton Mass
Limits", arXiv:0809.1003v3 [hep-ph]

\bibitem{Zurab}
  Z.~Kakushadze and P.~Langfelder,
  %``Gravitational Higgs mechanism,''
  Mod.\ Phys.\ Lett.\  A {\bf 15} (2000) 2265
  [arXiv:hep-th/0011245].
  %%CITATION = MPLAE,A15,2265;%%

\bibitem{jackiw} S.~Deser, R.~Jackiw and S.~Templeton,
  %``Topologically massive gauge theories,''
  Annals Phys.\  {\bf 140} (1982) 372
  [Erratum-ibid.\  {\bf 185} (1988\ APNYA,281,409-449.2000) 406.1988\ APNYA,281,409].
  %%CITATION = APNYA,281,409;%%

\bibitem{ma} E.~A.~Bergshoeff, O.~Hohm and P.~K.~Townsend,
  %``Massive Gravity in Three Dimensions,''
  Phys.\ Rev.\ Lett.\  {\bf 102} (2009) 201301
  [arXiv:0901.1766 [hep-th]].
  %%CITATION = PRLTA,102,201301;%%
  
\bibitem{ja} M.~Nakasone and I.~Oda,
  %``On Unitarity of Massive Gravity in Three Dimensions,''
  Prog.\ Theor.\ Phys.\  {\bf 121} (2009) 1389
  [arXiv:0902.3531 [hep-th]].
  %%CITATION = PTPKA,121,1389;%%

\bibitem{BMH} J. L. Boldo, L. M. de Moraes and J. A. Helay\"{e}l-Neto, Class.
Quantum Grav. 17 (2000) 813--823.

\bibitem{thooft} Gerard't Hooft "Unitarity in the Brout-Englert-Higgs
Mechanism for Gravity", arXiv:0708.3184v4 [hep-th].

\bibitem{Witten1} E. Witten, "Three-Dimensional Gravity Reconsidered,"
arXiv:0706.3359v1 [hep-th].

\bibitem{LSS} W. Li,W. Song and A. Strominger, JHEP 0804, 082 (2008)
[arXiv:0801.4566 [hep-th]].

\bibitem{utiyama} Ryoyu Utiyama, Phys. Rev. 101, 1597 - 1607 (1956).

\bibitem{kibble} T.~W.~B.~Kibble,
  %``Lorentz invariance and the gravitational field,''
  J.\ Math.\ Phys.\  {\bf 2} (1961) 212.
  %%CITATION = JMAPA,2,212;%%

\bibitem{bu} E. Sezgin and P. van Nieuwenhuizen, Phys.Rev.{\ D} \textbf{21},
3269 (1980).

\bibitem{bd} D. E. Neville, Phys. Rev.{\ D} \textbf{18}, 3535 (1978).
\end{thebibliography}
\end{document}